\documentclass[aps,prd,10pt,twocolumn,superscriptaddress,nofootinbib,floatfix]{revtex4-2}
\usepackage[utf8]{inputenc}
\usepackage{amsmath,amssymb}
\usepackage{graphicx}
\usepackage{booktabs}

\makeatletter
\newcommand{\raggedcells}{\expandafter\def\expandafter\@arrayparboxrestore\expandafter
 {\@arrayparboxrestore\raggedright\let\\\tabularnewline}}
\makeatother
\graphicspath{{figures/}{PRD_Letter/figures/}{./}}
\usepackage{microtype}
\usepackage{xcolor}
\usepackage[hidelinks]{hyperref}
\newcommand{\dd}{\mathrm d}
\newcommand{\mpl}{M_{\mathrm{Pl}}}
\newcommand{\Ncal}{\mathcal N}
\newcommand{\NGB}{\mathcal N_{\rm GB}}
\newcommand{\lth}{\lambda_{\rm tH}}

\begin{document}

\title{Gauss--Bonnet running and the de Sitter saddle of quadratic gravity inflation}

\author{Ruolin Liu}
\affiliation{Department of Physics and Astronomy, University of Waterloo, Waterloo, Ontario N2L 3G1, Canada}
\affiliation{Waterloo Centre for Astrophysics, University of Waterloo, Waterloo, Ontario N2L 3G1, Canada}
\affiliation{Perimeter Institute for Theoretical Physics, Waterloo, Ontario N2L 2Y5, Canada}
\author{Niayesh Afshordi}
\affiliation{Department of Physics and Astronomy, University of Waterloo, Waterloo, Ontario N2L 3G1, Canada}
\affiliation{Waterloo Centre for Astrophysics, University of Waterloo, Waterloo, Ontario N2L 3G1, Canada}
\affiliation{Perimeter Institute for Theoretical Physics, Waterloo, Ontario N2L 2Y5, Canada}

\begin{abstract}
Inflation driven purely by the quantum running of the curvature-squared
couplings of quadratic gravity was shown to accommodate the high
scalar tilt favored by recent cosmic microwave background observations,
starting from a Euclidean four-sphere at a maximum
of the running $R^2$ coupling. That maximum exists only in one renormalization
scheme, and is not stationary once the Euler trace
anomaly is included. We show that the standard one-loop running, with the
usually neglected Gauss--Bonnet coefficient retained, recovers the missing de
Sitter state: the Euler running balances the scale dependence of the $R^2$ term
at a unique coupling ratio, giving a de Sitter solution that is stationary for
the gravitational constraint and for the compact Euclidean action alike, and
whose scalaron potential is an extremely flat hilltop
($m^2/H^2\simeq-5\times10^{-10}$) joined to an inverse-linear inflationary
plateau. The last $N_*\simeq50$--$60$ $e$-folds are scheme independent, with
scalar tilt $n_s\simeq1-4/(3N_*)$, while the tensor-to-scalar ratio, $r$, is set by the
number of matter fields that enhance the running. The current tensor bound excludes pure gravity
and sets a minimum matter content, some $4\times10^6$ conformally coupled
scalars or $3\times10^5$ vectors---$2.6$ times fewer than the momentum-induced
scheme requires---while one-loop control to the end of inflation sets a maximum
about ten times higher. Across that window the model predicts $r\gtrsim0.008$ with
$0.973\lesssim n_s\lesssim0.978$, within reach of upcoming CMB polarization
surveys.
\end{abstract}

\maketitle

\textit{Introduction.---}Inflation can be driven by the gravitational field
itself. In the $R+R^2$ model the Einstein--Hilbert (EH) term shapes the
potential of the scalar curvature mode, the scalaron
\cite{Starobinsky:1980te,Stelle:1977ry}. Recent measurements have, however,
moved the scalar tilt above the Starobinsky prediction $n_s\simeq1-2/N_*$: the
combination of the Atacama Cosmology Telescope (ACT) DR6 spectra with Planck,
CMB lensing and DESI baryon acoustic oscillations gives $n_s=0.9743\pm0.0034$
\cite{ACT:2025lcdm,ACT:2025extended}, and the joint Planck--SPT--ACT--BICEP/Keck
analysis with DESI gives $0.9728\pm0.0029$ with $r<0.034$
\cite{Balkenhol:2025inf}.

Quadratic gravity without an EH term, i.e., with only curvature-squared terms,
behaves differently. It is
perturbatively renormalizable \cite{Stelle:1976gc} and its Weyl coupling is
asymptotically free \cite{Fradkin:1981iu,Avramidi:1985ki,Salvio:2014soa,Salvio:2018crh},
at the cost of a spin-2 ghost \cite{Stelle:1977ry}, the usual reason for
stopping at $R+R^2$. Here, however, the Weyl coupling becomes strong in the
infrared after inflation, where the ghost may be confined and the EH term
generated, as in the QCD analogy of Holdom and Ren
\cite{Holdom:2015kbf,Holdom:2016xfn}. With constant couplings and no EH term
the scalaron potential is flat and
the classical theory has neither a curvature scale nor a slope that would end
inflation \cite{Alvarez-Gaume:2015rwa}. Quantum running lifts this degeneracy:
relating the renormalization scale to curvature turns the beta functions into a
potential \cite{Platania:2020lqb}. In Ref.~\cite{Liu:2025qg} we showed that
momentum-induced beta functions \cite{Buccio:2024hys} on an asymptotically free
trajectory give a viable inflationary scenario in which many matter fields lower
the tensor amplitude. Two questions remained. Does the mechanism depend on the
maximum of the $R^2$ coupling that exists for those beta functions but
not for the standard local ones, and what plays the role of the early de Sitter
(dS) state? The second is sharpened by an
observation of Shinn \cite{Shinn:2026pc}: on a round four-sphere the Euler
($a$-type) trace anomaly contributes to the scale derivative of the action
independently of the RG scheme, so the sphere of Ref.~\cite{Liu:2025qg}, placed
where $\beta_\xi=0$, is not stationary once that contribution is kept.

Here we answer both within the standard local one-loop running, keeping all
three couplings. The third, the coefficient of the Gauss--Bonnet or Euler
density, is topological when constant and is usually dropped; its running is
nevertheless part of the scale dependence of the action
\cite{Einhorn:2014abc}, and, as in classically scale-invariant gravity
\cite{Einhorn:2014gfa,Hawking:2000bb}, it selects a nonzero-curvature extremum.
We show that this extremum is a dS saddle of the RG-improved theory, place it on
the same trajectory that later produces the observed perturbations, derive
closed-form observables for the rolling regime, and confront them with
current data. Throughout, running couplings are inserted as functions of the
curvature before the metric is varied (a local RG improvement), signature is
$(-,+,+,+)$ and $\hbar=c=1$.

\textit{Running couplings.---}The action is
\begin{equation}
 S=\int\!\dd^4x\sqrt{-g}\Big[-\frac{R^2}{\xi}-\frac{C_{\mu\nu\rho\sigma}C^{\mu\nu\rho\sigma}}{2\lambda}+\sigma E_4\Big],
 \label{eq:action}
\end{equation}
with $C$ the Weyl tensor, $E_4=R_{\mu\nu\rho\sigma}^2-4R_{\mu\nu}^2+R^2$ and no EH
or cosmological terms. We study the branch $\lambda>0$, $\xi<0$, on which the
$R^2$ coefficient is positive and the scalaron is not tachyonic at fixed
couplings. Writing
$t$ for the logarithm of the renormalization scale, with an origin fixed below,
and $\beta_q=\dd q/\dd t$, the standard one-loop flow with
$N_V$ vectors, $N_F$ Weyl fermions and $N_S$ conformally coupled real scalars is
\cite{Fradkin:1981iu,Avramidi:1985ki,Salvio:2014soa}
\begin{align}
 16\pi^2\beta_\lambda&=-\Ncal\lambda^2,\qquad
 16\pi^2\beta_\sigma=\NGB,\nonumber\\
 16\pi^2\beta_\xi&=-10\lambda^2+5\lambda\xi-\tfrac{5}{36}\xi^2,
 \label{eq:betas}\\
 \Ncal&=\tfrac{133}{10}+\tfrac{N_V}{5}+\tfrac{N_F}{20}+\tfrac{N_S}{60},\nonumber\\
 \NGB&=\tfrac{196}{45}+\tfrac{1}{360}\big(62N_V+\tfrac{11}{2}N_F+N_S\big).\nonumber
\end{align}
Thus $\Ncal$ and $\NGB$ are total coefficients, including the gravitational
parts, rather than species counts, and non-negative counts confine $\NGB$ to
$\frac{196}{45}+\frac{\Ncal-13.3}{6}\le\NGB\le\frac{196}{45}+\frac{31(\Ncal-13.3)}{36}$
(scalar-only and vector-only edges); Ref.~\cite{Liu:2025qg} denotes by
$\Ncal$ the matter part alone, which differs from ours by $133/10$. The matter
fields are free and massless, so they enter only through the coefficients of
Eq.~\eqref{eq:betas}, without thresholds or self-interactions.

We fix the RG origin at the formal crossing $\xi(0)=0$, $\lambda(0)=\lambda_0$,
and define the 't Hooft-like coupling $\lth\equiv\Ncal\lambda_0/16\pi^2$. At this origin the $R^2$
coupling $f_0^2\equiv-\xi/6$ vanishes and the $R^2$ coefficient of
Eq.~\eqref{eq:action} diverges, but the history considered here stays at
$t>0$. Every term in $\beta_\xi$ is negative on our
branch, so $\xi$ decreases monotonically and the $R^2$ coupling alone never has a
stationary point. The flow is nevertheless exactly solvable: with
$\gamma\equiv\lambda/\xi$, $\gamma_\pm^{\rm RG}=[\Ncal+5\pm\sqrt{(\Ncal+5)^2-50/9}]/20$
and $\nu=\sqrt{(\Ncal+5)^2-50/9}/\Ncal$,
\begin{equation}
 \lambda=\frac{\lambda_0}{1+\lth t},\qquad
 \xi=\lambda\,\frac{(1+\lth t)^\nu-1}{\gamma_-^{\rm RG}(1+\lth t)^\nu-\gamma_+^{\rm RG}},
 \label{eq:orbit}
\end{equation}
and $\sigma=\sigma_0+\NGB\,t/16\pi^2$.
The negative-$\xi$ branch starts at $t=0^+$ and ends at the Riccati pole
$(1+\lth t)^\nu=\gamma_+^{\rm RG}/\gamma_-^{\rm RG}$.

\textit{The Euler-selected de Sitter saddle.---}A maximally symmetric geometry
has $C=0$ and $E_4=R^2/6$, so on such backgrounds only the combination
\begin{equation}
 h\equiv\frac{\sigma}{6}-\frac1\xi=\frac{C_E}{6}+\frac{1}{\xi'},\qquad
 \frac1{\xi'}\equiv\frac{\NGB}{6\Ncal\lambda}-\frac1\xi,
 \label{eq:hCE}
\end{equation}
enters the action, where $C_E\equiv\sigma-\NGB/(\Ncal\lambda)$ is the additive
integration constant of $\sigma$ and is exactly conserved by Eq.~\eqref{eq:betas}.
Being the coefficient of a topological term, $C_E$ drops out of every local
equation; we therefore build the Euler-completed coupling $\xi'$ from the running
part of $\sigma$ alone, so that $\xi'$ is a function of $(\lambda,\xi)$, is
positive on our branch, obeys $\dd h/\dd t=\dd(1/\xi')/\dd t$, and closes with
$\lambda$ into a two-dimensional flow.

\begin{figure*}[t]
 \centering
 \includegraphics[width=0.98\textwidth]{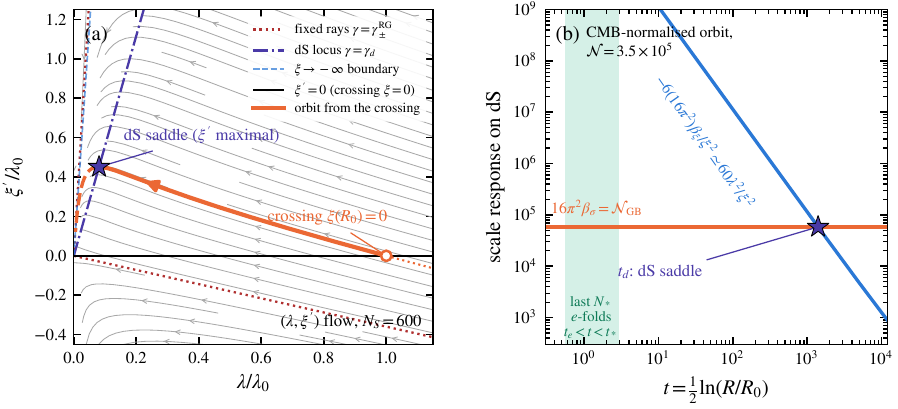}
 \caption{(a) Renormalization-group flow in the Euler-completed
 coordinates $(\lambda,\xi')$ of Eq.~\eqref{eq:hCE}, in units of the crossing
 value $\lambda_0$, for $N_S=600$ ($\Ncal=23.3$, $\NGB=6.02$); arrows point
 toward increasing $\mu$. The orange curve is the exact orbit \eqref{eq:orbit} from the
 crossing (open circle; dotted: formal positive-$\xi$ continuation). It meets the
 negative-$\xi$ dS locus $\gamma=\gamma_d$ (dash-dotted) at the star, where $\xi'$
 is maximal, $\lambda_d/\lambda_0=0.081$, $\xi'_d/\lambda_0=0.45$, and continues
 (dashed) toward the $\xi\to-\infty$ boundary. The black line $\xi'=0$ is the
 crossing locus $\xi=0$; dotted red lines are the fixed rays
 $\gamma=\gamma^{\rm RG}_\pm$. (b) The two terms of the dS condition
 \eqref{eq:dScond} along the CMB-normalized orbit ($N_*=55$,
 $\lth=0.81$): the $R^2$ response falls as $60\lambda^2/\xi^2\propto t^{-2}$
 while the Euler response is constant; they balance at $t_d\simeq1.4\times10^3$
 (star). The shaded band is the observable interval $t_e<t<t_*$.}
 \label{fig:flow}
\end{figure*}

The scale dependence of the action contributes to the trace of the
gravitational equations $\Theta_\beta=(\beta_\xi/\xi^2)R^2+(\beta_\lambda/2\lambda^2)C^2+\beta_\sigma E_4$
(modulo $\Box R$ and equation-of-motion terms \cite{Duff:1993wm,Einhorn:2014gfa}).
We identify $\mu^2\propto R$, i.e.\ $t=\frac12\ln(R/R_0)$, and vary the
RG-improved action $L(R,G)=-R^2/\xi(R)+\sigma(R)G$, $G=E_4$, on spatially flat
FLRW. The lapse variation gives the homogeneous constraint
$E_0=-L+GL_G+6(\dot H+H^2)L_R-6H\dot L_R-24H^3\dot L_G=0$, where dots are
cosmic-time derivatives (Appendix~\ref{app:constraint}). On dS with constant $H$ ($R=12H^2$,
$E_4=24H^4$, $C^2=\Box R=0$) both routes reduce to
\begin{equation}
 E_0\big|_{\rm dS}=\frac{R^2}{4}\Big(\frac{\beta_\xi}{\xi^2}+\frac{\beta_\sigma}{6}\Big)=0
 \;\Longleftrightarrow\;\frac{\dd h}{\dd t}=0 .
 \label{eq:dScond}
\end{equation}
The $R^2$ and Euler responses cancel although neither coupling is stationary
[Fig.~\ref{fig:flow}(b)]. With Eq.~\eqref{eq:betas} this fixes the coupling
ratio,
\begin{align}
 D_{\rm dS}(\gamma)&\equiv-360\gamma^2+180\gamma+6\NGB-5=0,\nonumber\\
 \gamma_d&=\frac14-\frac{\sqrt{175+60\NGB}}{60},
 \label{eq:gammad}
\end{align}
where the negative root lies on our branch whenever $\NGB>5/6$, which holds for
any matter content. Without Euler running, $\beta_\xi<0$ would leave no finite
root: the Gauss--Bonnet coefficient is what selects the dS solution. In vacuum,
Eq.~\eqref{eq:gammad} reproduces the compact-curvature extremum of
Ref.~\cite{Einhorn:2014gfa}.

Since $\dd h/\dd t=D_{\rm dS}(\gamma)/(36\cdot16\pi^2)$ and
$16\pi^2\beta_\gamma=\lambda[10\gamma^2-(\Ncal+5)\gamma+5/36]>0$ for $\gamma<0$,
one finds $\dd^2\xi'/\dd t^2|_d=-\frac{5}{16\pi^2}\xi'^2_d(1-4\gamma_d)\beta_{\gamma,d}<0$:
the Euler-completed coupling $\xi'$ has a strict maximum where $\xi$ has none
[Fig.~\ref{fig:flow}(a)]. This is not an RG fixed point ($\beta_\lambda\neq0$
there) but a stationary point of the RG-improved cosmological action. The same
condition extremizes the Euclidean action of the round four-sphere and of the
hemisphere, $I_{E,1/2}(t)=192\pi^2/\xi-32\pi^2\sigma=-192\pi^2/\xi'-32\pi^2C_E$,
with $I_{E,1/2,tt}|_d=-192\pi^2h_{tt}|_d<0$: the hemisphere is a regular candidate
saddle of a no-boundary amplitude
\cite{Hartle:1983ai,Hawking:1984ph,Lehners:2023yrj}, and $C_E$ enters only as a
constant shift of its action (Appendix~\ref{app:euclid}).

On the crossing-normalized orbit \eqref{eq:orbit} the saddle sits at a unique
RG time,
\begin{equation}
 \lambda_d=\lambda_0\Big(\frac{\gamma_d-\gamma_-^{\rm RG}}{\gamma_d-\gamma_+^{\rm RG}}\Big)^{1/\nu},\quad
 t_d=\frac{\lambda_0/\lambda_d-1}{\lth},\quad \xi_d=\frac{\lambda_d}{\gamma_d},
 \label{eq:saddle}
\end{equation}
with $0<t_d<t_{\rm pole}$. For the CMB-normalized trajectory found below
($\Ncal=3.5\times10^5$, scalar matter) $\gamma_d=-30.9$,
$\lambda_d=3.2\times10^{-7}$, $\xi_d=-1.0\times10^{-8}$ and $t_d\simeq1.4\times10^3$
[Fig.~\ref{fig:flow}(b)], while the pole is at $t\simeq10^{11}$; $\lambda_d$ and $|\xi_d|$ are tiny at the saddle.

\begin{figure}[t]
 \centering
 \includegraphics[width=\linewidth]{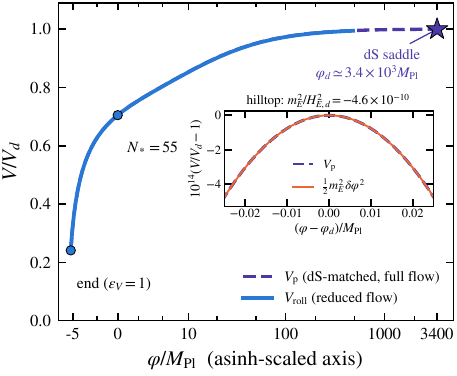}
 \caption{Einstein-frame scalaron potential on the CMB-normalized trajectory,
 from the end of inflation to the dS saddle, on an asinh-scaled field axis
 ($R_0/\mpl^2$ chosen so that $\varphi_*=0$). Dashed: the dS-matched potential
 \eqref{eq:Vp} evaluated on the full orbit \eqref{eq:orbit}; solid: the reduced
 rolling potential \eqref{eq:Vroll}, drawn up to $\varphi\simeq5\times10^2\mpl$.
 The two agree in height to better than $10^{-3}$ over the whole range shown
 and in canonical slope to $1.3\times10^{-3}$ in the observable window (at
 equal RG time); only the full potential has the maximum. Inset: the hilltop \eqref{eq:hilltop} against $V_{\rm p}$.}
 \label{fig:potential}
\end{figure}

\textit{A shallow hilltop.---}To describe departures from the saddle we use the
scalaron. With running $\sigma$, $L(R,G)$ is not an $f(R)$ theory (its Hessian
in $(R,G)$ is nondegenerate), but near dS an $f(R)$ action with
the same quadratic dynamics exists. Subtracting the constant $\sigma_d=\sigma(R_d)$
and defining $c_{\rm p}(R)=-1/\xi+(\sigma-\sigma_d)/6=1/\xi'-\NGB/(6\Ncal\lambda_d)$, the
identity $-R^2/\xi+\sigma E_4=R^2c_{\rm p}+\sigma_dE_4+(\sigma-\sigma_d)(E_4-R^2/6)$
isolates a remainder that is cubic in perturbations about dS, since
$E_4-R^2/6=C^2-2S_{\mu\nu}S^{\mu\nu}$ with $S_{\mu\nu}$ the traceless Ricci
tensor (Appendix~\ref{app:scalaron}). Hence $f_{\rm p}(R)=R^2c_{\rm p}(R)$ reproduces the theory to
quadratic order. Its Einstein-frame potential and canonical field are, with
$x=\ln(R/R_d)$ and $F=f_{{\rm p},R}=R(2c_{\rm p}+c_{{\rm p},x})$
\cite{DeFelice:2010aj},
\begin{equation}
 V_{\rm p}=\frac{\mpl^4}{4}\frac{c_{\rm p}+c_{{\rm p},x}}{(2c_{\rm p}+c_{{\rm p},x})^2},\qquad
 \varphi=\sqrt{\tfrac32}\,\mpl\ln\frac{2F}{\mpl^2},
 \label{eq:Vp}
\end{equation}
where $\mpl$ normalizes the Einstein-frame gravitational term and is not an EH
coupling of Eq.~\eqref{eq:action}. Because
$c_{{\rm p},x}=\frac12(\beta_\xi/\xi^2+\beta_\sigma/6)$, the potential is
stationary exactly where Eq.~\eqref{eq:dScond} holds, at height
$V_d=-\mpl^4\lambda_d/16\gamma_d$, and its curvature,
$m_E^2\equiv V_{{\rm p},\varphi\varphi}|_d$, is fixed by the running:
\begin{equation}
 \frac{m_E^2}{H_{E,d}^2}=-4\Big[1-\frac{8(16\pi^2)^2\gamma_d}
 {5\lambda_d^2(1-4\gamma_d)(\NGB/6-\Ncal\gamma_d)}\Big]^{-1}<0,
 \label{eq:hilltop}
\end{equation}
with $H_{E,d}^2=V_d/3\mpl^2$. The saddle is a local maximum of the scalaron
potential, but an extraordinarily flat one: for the CMB-normalized trajectory
$m_E^2/H_{E,d}^2=-4.6\times10^{-10}$ (Fig.~\ref{fig:potential}, inset), so
that the classical roll-off time, $3H_{E,d}/|m_E^2|\sim10^{10}$ Hubble times, is far
longer than the quantum diffusion time. Exact dS remains a solution; the
instability supplies a direction of departure, not its initial amplitude, which
must come from the quantum state \cite{Lehners:2023yrj,Maldacena:2024uhs}.

\textit{The rolling regime.---}Cosmological rolling toward lower curvature
proceeds toward smaller $t$. Far below the saddle, $|\xi/\lambda|\ll1$ and the
$\lambda\xi$, $\xi^2$ and Euler terms are subleading; the leading flow
$16\pi^2\beta_\lambda=-\Ncal\lambda^2$, $16\pi^2\beta_\xi=A\lambda^2$ with
$A=-10$ integrates to
\begin{equation}
 \xi(t)=\frac{A\lambda_0^2}{16\pi^2}\frac{t}{1+\lth t},\qquad
 \frac{\xi}{\lambda}=\frac{A\lambda_0 t}{16\pi^2}.
 \label{eq:reduced}
\end{equation}
Here $A$ labels the beta-function prescription rather than a fit parameter.
For $f(R)=-R^2/\xi(R)$ with $R=R_0e^{2t}$ and the abbreviation
$D(t)\equiv t(1+\lth t)$, the Einstein-frame potential and field are
\begin{align}
 V_{\rm roll}(t)&=\frac{\mpl^4|A|\lambda_0^2}{16\pi^2}\,\frac{t^2\,[2D-1]}{2\,[4D-1]^2},
 \label{eq:Vroll}\\
 \frac{\varphi(t)}{\mpl}&=\sqrt{\tfrac32}\ln\Big[\frac{16\pi^2R_0}{|A|\lambda_0^2\mpl^2}\frac{e^{2t}(4D-1)}{t^2}\Big].
 \nonumber
\end{align}
The chart requires $D>1/4$ and $V>0$ requires $D>1/2$; the crossing $t=0$ is
never reached by the scalaron description. For $\lth t\gg1$,
\begin{equation}
 V_{\rm roll}\simeq V_0\Big[1-\frac{\sqrt6\,\mpl}{\lth(\varphi-\bar\varphi_0)}\Big],\qquad
 V_0=\frac{\mpl^4|A|\lambda_0}{16\Ncal},
 \label{eq:plateau}
\end{equation}
which is exactly the inverse-linear plateau of Ref.~\cite{Liu:2025qg}, with constant offset
$\bar\varphi_0$ (Appendix~\ref{app:roll}). For the trajectory of Fig.~\ref{fig:potential},
$V_0/V_d=1.0008$: the plateau and the shallow hilltop at
$\varphi_d-\varphi_*\simeq\sqrt6\,(t_d-t_*)\mpl\simeq3.4\times10^3\mpl$ are the
two ends of one potential of almost constant height. The plateau is formally
infinite only in the reduced flow; on the full orbit $|\xi/\lambda|$ grows and
the Euler-selected saddle terminates it at finite curvature. Since
$N\simeq2\lth t^3$ along the plateau, a field released anywhere between the
saddle and the pivot inflates for an enormous number of $e$-folds before
entering the observable window, which is therefore insensitive to the initial
condition.

\textit{Observables.---}Canonical differentiation of Eq.~\eqref{eq:Vroll}
gives $\mpl V_{,\varphi}/V=\sqrt{2/3}/(2D-1)$ exactly, so with $D_*=D(t_*)$
\begin{align}
 r&=\frac{16}{3(2D_*-1)^2},\nonumber\\
 n_s&=1-\frac{2}{(2D_*-1)^2}-\frac{4[1+\lth t_*(1+4D_*)]}{3(2D_*-1)P_*},
 \label{eq:nsr}
\end{align}
with $P_*\equiv4D_*^2-3D_*+1+\lth t_*$, at first order in slow roll, and the
scalar amplitude is
\begin{equation}
 A_s=\frac{|A|\lth^2}{\Ncal^2}\,\frac{t_*^2(2D_*-1)^3}{(4D_*-1)^2}.
 \label{eq:As}
\end{equation}
Inflation ends at $\epsilon_V\equiv\tfrac12\mpl^2(V_{,\varphi}/V)^2=1$, i.e.\ $2D(t_e)-1=1/\sqrt3$, and the number of
Einstein-frame $e$-folds follows from the closed primitive
\begin{equation}
 I(t)=2\lth t^3+3t^2-6t+3\ln t-\tfrac34\ln(4D-1),
 \label{eq:Nefold}
\end{equation}
$N_*=I(t_*)-I(t_e)$.
Given $N_*$ and $\lth$, Eq.~\eqref{eq:Nefold} fixes $t_*$, Eq.~\eqref{eq:nsr}
the observables, and Eq.~\eqref{eq:As} the coefficient $\Ncal$ (hence
$\lambda_0=16\pi^2\lth/\Ncal$). No independent pivot coupling is introduced.
Two limits follow. For $\lth t_*\ll1$ (pure or weakly enhanced running)
$N_*\simeq3t_*^2$, $n_s\simeq1-3/(2N_*)$ and $r\simeq4/N_*$. For $\lth t_*\gg1$
(matter-enhanced running)
\begin{equation}
 n_s\simeq1-\frac{4}{3N_*},\quad
 r\simeq\frac83\Big(\frac{2}{\lth^2N_*^4}\Big)^{1/3}
 =\frac43\Big(\frac{2|A|}{A_s\Ncal^2N_*^2}\Big)^{2/5},
 \label{eq:limits}
\end{equation}
where the two forms of $r$ hold $\lth$ or $\Ncal$ fixed and the first is that
of Ref.~\cite{Liu:2025qg}; the tilt is set by the duration alone while the
tensor amplitude is set by the matter content. The leading predictions depend on the prescription only
through $A_s\Ncal^2/|A|$, with $A=-10$ here and $A=70$ for the momentum-induced
flow \cite{Buccio:2024hys}, which is the universality noted in
Ref.~\cite{Liu:2025qg}; the figures use the unexpanded expressions.

The theory has the constants
$(\lambda_0,\Ncal,\NGB,C_E)$ and the cosmology adds $N_*$. Because a constant
Euler coefficient is topological, $C_E$ drops out of the constraint, of the
potential and of every observable: it is exactly degenerate and is not a fit
parameter, entering only the Euclidean weight $e^{32\pi^2C_E\chi}$ of a given
topology. $\NGB$ enters only through the saddle and the negligible correction
estimated below. Once Eq.~\eqref{eq:As} fixes $\lambda_0$, the observables at
given $N_*$ depend on the single combination $\lth$.

\begin{figure*}[t]
 \centering
 \includegraphics[width=0.98\textwidth]{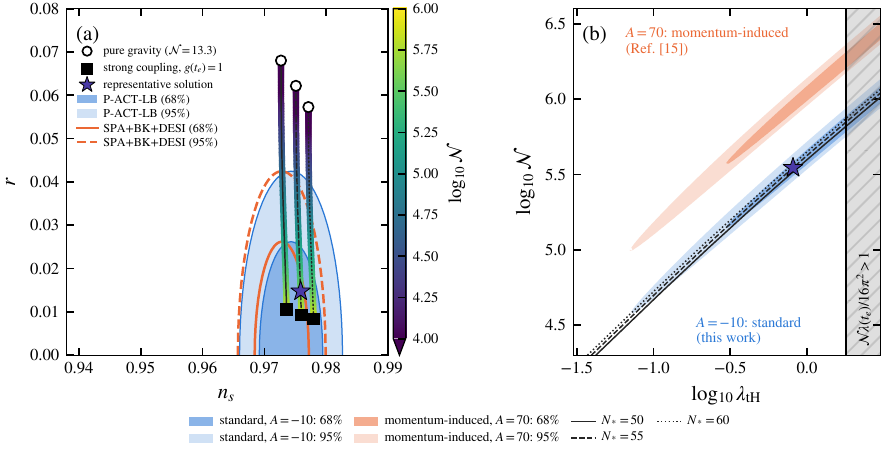}
 \caption{(a) Predictions of Eqs.~\eqref{eq:nsr}--\eqref{eq:Nefold} for
 $N_*=50,55,60$ as $\lth$ runs from the pure-gravity value (open circles,
 $\Ncal=13.3$) to the strong-coupling limit $g(t_e)\equiv\Ncal\lambda(t_e)/16\pi^2=1$
 (filled squares), colored by $\log_{10}\Ncal$ from Eq.~\eqref{eq:As} with
 $A_s=2.1\times10^{-9}$; overlaid solid, dashed and dotted lines mark
 $N_*=50,55,60$ in both panels. Shaded: 68\% and 95\% regions of the approximate
 likelihood $\chi^2=[(n_s-0.9743)/0.0034]^2+(r/0.0173)^2$ (P-ACT-LB tilt
 \cite{ACT:2025lcdm}; half-Gaussian in $r\ge0$ whose 95\% quantile is the
 measured limit $r<0.034$ \cite{Balkenhol:2025inf}); orange: the same with
 $n_s=0.9728\pm0.0029$ (Planck--SPT--ACT--BK--DESI \cite{Balkenhol:2025inf}).
 At fixed $(\lth,N_*)$ these curves coincide with those of the momentum-induced flow of Ref.~\cite{Liu:2025qg}. (b) The same approximate
 likelihood mapped to the plane of the enhanced crossing coupling $\lth$ and
 total coefficient $\Ncal$, with $t_*$ fixed by $A_s$ and $N_*$ left free, for
 the standard flow ($A=-10$, blue) and the momentum-induced flow of
 Refs.~\cite{Buccio:2024hys,Liu:2025qg} ($A=70$, orange); darker and lighter
 shading are the 68\% and 95\% regions, and the dark lines locate the curves
 of (a). The bands are parallel,
 $\Ncal\propto\sqrt{|A|}$ at fixed $(\lth,N_*)$, and both are cut by the
 strong-coupling limit (hatched).}
 \label{fig:nsr}
\end{figure*}

\textit{Confrontation with data.---}Figure~\ref{fig:nsr} shows the prediction
curves against the 2025 constraints. Pure gravity ($N_S=N_F=N_V=0$,
$\Ncal=13.3$, $\lth\simeq2.6\times10^{-5}$) gives $n_s=0.973$--$0.977$ but
$r=0.057$--$0.068$ for $N_*=50$--$60$, excluded at $3.3$--$3.9\sigma$ by the
BICEP/Keck-based bound \cite{BICEP:2021xfz,Tristram:2021tvh,Balkenhol:2025inf}.
Matter-enhanced running is therefore required: $r<0.034$ demands
$\lth>0.11$--$0.19$, i.e.\ $\Ncal>(6.0$--$8.8)\times10^4$, or
$N_S\gtrsim4\times10^6$ conformal scalars (equivalently $N_V\gtrsim3\times10^5$
vectors). The loop-expansion
parameter of the matter-dominated running is
$g(t)\equiv\Ncal\lambda(t)/16\pi^2=\lth/(1+\lth t)$, which increases toward the
end of inflation; requiring $g(t_e)<1$ gives $\lth<1.79$ and hence a minimum
tensor amplitude,
\begin{equation}
 r_{\min}=0.0105,\;0.0093,\;0.0084\quad\text{for}\quad N_*=50,\,55,\,60,
 \label{eq:rmin}
\end{equation}
at $\Ncal=(6.6$--$7.4)\times10^5$, i.e.\ $N_S\simeq4\times10^7$ scalars or
$N_V\simeq3.5\times10^6$ vectors, in line with the $r\gtrsim0.01$ of
Ref.~\cite{Liu:2025qg}. The viable matter content is thus
$4\times10^6\lesssim N_S\lesssim4\times10^7$ (or $3\times10^5\lesssim N_V\lesssim4\times10^6$). Across the allowed window the tilt barely moves,
$n_s=0.9730$--$0.9736$, $0.9754$--$0.9760$ and $0.9773$--$0.9779$ for
$N_*=50,55,60$, tracking $1-4/(3N_*)$. Relative to the ACT tilt the
model is an excellent fit ($\chi^2\lesssim1.4$ for all $N_*$ at the strong-coupling
edge); relative to the Planck--SPT--ACT--BK--DESI value it prefers $N_*\lesssim55$,
and the combination without DESI, $n_s=0.9682\pm0.0032$
\cite{Balkenhol:2025inf}, is in $2.2$--$3\sigma$ tension for $N_*\ge55$.

A representative solution, used for Figs.~\ref{fig:flow}(b) and
\ref{fig:potential}, has $N_*=55$, $\lth=0.810$ ($D_*=10$), $t_*=2.950$,
$t_e=0.547$, $\Ncal=3.50\times10^5$, $\lambda_0=3.65\times10^{-4}$, and predicts
$n_s=0.9759$, $r=0.0148$; its couplings are $g(t_*)=0.24$ and $g(t_e)=0.56$.
Neither a small individual coupling nor $g<1$ is by itself a bound on neglected
higher loops \cite{Salvio:2017qkx}, and the required matter content
($N_S\simeq2\times10^7$) is large, as in the momentum-induced scenario.

\textit{Relation to the momentum-induced scenario.---}The earlier
analysis~\cite{Liu:2025qg} used the same action, identification $\mu^2\propto R$ and
crossing normalization, but the momentum-induced beta functions
\cite{Buccio:2024hys}, in which $16\pi^2\beta_\xi=70\lambda^2+\ldots$ and
$16\pi^2\beta_\lambda=-(\Ncal+\frac{14}{3})\lambda^2+\ldots$ at fixed species
content, and no Euler term. Three differences follow. First, there $\beta_\xi$
has a zero on the physical branch, and the maximum of the $R^2$ coupling was
taken to select the Euclidean four-sphere. On $S^4$, however,
$\dd I_E/\dd t=-384\pi^2\beta_\xi/\xi^2-4\NGB$, the second term being the
scheme-independent Euler anomaly \cite{Shinn:2026pc}; with the momentum-induced
$\beta_\xi$ the corrected condition,
$(1-6\NGB)(\xi/\lambda)^2-36\,\xi/\lambda-2520=0$, has no real root for
$\NGB>79/420$, i.e.\ already for pure gravity, so that flow has no stationary
round sphere at all. Here $\xi$ is monotonic and the sphere is selected by the
Euler balance \eqref{eq:dScond}, which has a root on our branch for every
$\NGB>5/6$. Second, there both couplings are
asymptotically free toward the UV, whereas here $\lambda$ is asymptotically
free but $\xi$ reaches the pole $f_0^2\to\infty$ \cite{Salvio:2017qkx} at
$t_{\rm pole}\gg t_d$; the saddle, not the pole, is the highest-curvature
state of the cosmological history. Third,
the rolling regimes coincide: both flows reduce to Eq.~\eqref{eq:reduced},
the potential \eqref{eq:plateau} is the inverse-linear plateau of
Ref.~\cite{Liu:2025qg} with $|A|=70$, and $n_s$, $r$ and $r_{\min}$ are
unchanged at fixed $(\lth,N_*)$, while the matter content required by $A_s$
scales as $\Ncal\propto\sqrt{|A|}$, so the standard prescription needs
$\sqrt7\simeq2.6$ times fewer fields [Fig.~\ref{fig:nsr}(b)]. The
inflationary predictions are thus scheme independent, but the early dS state
is not: in the standard scheme it exists only because of the running
Gauss--Bonnet term (Appendix~\ref{app:compare}).

\textit{Euler and Weyl corrections.---}The reduced potential omits the running
Euler term during rolling and the Weyl term, which vanishes on FLRW but affects
tensors. Projecting the Euler coupling onto an Einstein-scalar--Gauss--Bonnet
form, $-\Xi(\varphi)E_4/2$ with $\Xi=-2\sigma$, the leading correction to the
scalar equation, $\dd\varphi/\dd\ln a\simeq-\mpl^2V_{,\varphi}/V\,(1+d)$
\cite{Guo:2010jr}, has
$d=-2\NGB A_sD_*(4D_*-1)/[3(2D_*-1)P_*]$
(Appendix~\ref{app:corrections}), which for the representative solution is
$|d|=4.5\times10^{-6}$--$2.3\times10^{-5}$ across the allowed $\NGB$ range;
the corresponding shift of the tensor spectrum is below $10^{-7}$. The smallness
of $A_s$ and the finite range of enhanced running make the Euler term negligible
in the observable window, even though it balances the curvature response at the
much earlier saddle. The spin-2 mode of the Weyl sector has
$M_2^2/H_E^2=3\Ncal(4D-1)^2/[10\lth t\,D(2D-1)]\simeq3.5\times10^5$ at the
pivot, so its heavy-mode correction to the tensor normalization,
$(1+2H_E^2/M_2^2)^{-1}$, differs from unity by $6\times10^{-6}$
\cite{Clunan:2009er,Deruelle:2010kf,Salvio:2017xul}; the interpretation of the
associated negative-residue pole remains a quantization issue
\cite{Donoghue:2018izj,Anselmi:2020lpp}.

\textit{Discussion.---}The standard one-loop running of quadratic gravity,
with the Gauss--Bonnet coefficient retained, contains a complete inflationary
scenario on a single RG trajectory. At high curvature the Euler running selects
a dS saddle at which the constraint, the Euler-completed coupling and the
compact Euclidean action are all stationary; the scalaron potential there is a
hilltop so flat that its departure is quantum dominated. The hilltop continues
into an inverse-linear plateau, and the last $N_*$ $e$-folds give closed-form
observables normalized at the formal
crossing $\xi(R_0)=0$ rather than at an independently chosen pivot. The dS
saddle therefore does not replace an asymptotically free UV completion; rather,
it supplies the early stationary state whose absence was noted for the
standard beta functions.

The tilt is fixed by the duration,
$n_s\simeq1-4/(3N_*)$, above the Starobinsky value and in agreement with the
ACT-based measurements; the tensor amplitude is confined to
$0.008\lesssim r\lesssim0.034$ by the current bound from above and by one-loop
control to the end of inflation from below. LiteBIRD, with
$\sigma(r)\sim10^{-3}$ \cite{LiteBIRD:2022cnt}, will either
detect the tensor signal or exclude the mechanism, and CMB-HD, with
$\sigma(n_s)=0.0013$ \cite{MacInnis:2023vif}, would resolve the $N_*$
dependence of the tilt. Open issues are the
generation of the EH term and reheating after $t_e$, where the Einstein-frame
potential vanishes at $D=1/2$ and the running becomes strong
($g\to1$ at $t=1-1/\lth$); higher-loop control of the large matter sector;
the full perturbation theory of the running-Euler action beyond the two
descriptions compared in Fig.~\ref{fig:potential}; and the quantum
state that prepares the departure from the saddle. The appendices contain the derivations, the comparison with momentum-induced beta functions, and a description of the code that produces all figures, which is provided as ancillary files.

\begin{acknowledgments}
We thank M.~Shinn for pointing out that the four-sphere of
Ref.~\cite{Liu:2025qg} is not stationary once the Euler anomaly is retained,
and J.~Quintin, H.~Jiang and Y.~Manita for discussions. This research was
supported in part by Perimeter Institute for Theoretical Physics. Research at
Perimeter Institute is supported in part by the Government of Canada through the Department of Innovation,
Science and Economic Development and by the Province of Ontario through the
Ministry of Colleges and Universities. R.L. and N.A. are supported by the
Natural Sciences and Engineering Research Council of Canada (NSERC).
Claude 5.1 and ChatGPT 6 assisted with writing and cross-verifying the
derivations and accompanying code; the authors verified all results and take
full responsibility for the contents.
\end{acknowledgments}

\appendix
\onecolumngrid

\section{Conventions and dictionary}

The action \eqref{eq:action} is related to the ``agravity'' parametrization
of Refs.~\cite{Salvio:2014soa,Salvio:2018crh},
$\mathcal L=\frac{R^2}{6f_0^2}+\frac{\frac13R^2-R_{\mu\nu}^2}{f_2^2}$, by the
Gauss--Bonnet identity $C^2=E_4-2(\frac13R^2-R_{\mu\nu}^2)$:
\begin{equation}
 \lambda=f_2^2,\qquad \xi=-6f_0^2,\qquad
 -\frac{C^2}{2\lambda}=-\frac{E_4}{2\lambda}+\frac{\frac13R^2-R_{\mu\nu}^2}{\lambda}.
\end{equation}
The one-loop beta functions of $f_2^2$ and $f_0^2$ with conformally coupled
scalars, $(4\pi)^2\beta_{f_2^2}=-\Ncal f_2^4$ and
$(4\pi)^2\beta_{f_0^2}=\frac53f_2^4+5f_2^2f_0^2+\frac56f_0^4$
\cite{Fradkin:1981iu,Avramidi:1985ki,Salvio:2014soa}, translate into
Eq.~\eqref{eq:betas}. The branch $\lambda>0$, $\xi<0$ is $f_2^2>0$, $f_0^2>0$,
on which neither the scalaron nor the Weyl mode is tachyonic at fixed couplings;
the small running-induced curvature of the scalaron potential at the dS saddle
is computed in Appendix~\ref{sec:scalaron}. The Euler
coefficient $\NGB=\frac{196}{45}+\frac{1}{360}(62N_V+\frac{11}{2}N_F+N_S)$
combines the gravitational contribution of quadratic gravity
\cite{Fradkin:1981iu,Avramidi:1985ki} with the standard $a$-type trace-anomaly
coefficients of free conformal matter; a Dirac fermion counts as two Weyl
fermions. Throughout, $t=\frac12\ln(R/R_0)$ is the RG time, a subscript variable
(with or without a comma) denotes a partial derivative, $L_R=\partial L/\partial R$,
$h_{tt}=\dd^2h/\dd t^2$, $V_{,\varphi}=\dd V/\dd\varphi$, and $\xi'$ is the
Euler-completed coupling, not a derivative.

\section{Homogeneous constraint of the RG-improved action}
\label{app:constraint}\label{sec:constraint}

Insert the running couplings at $\mu^2\propto R$ and write the Lagrangian on
spatially flat FLRW as $L(R,G)$ with $G=E_4=24H^2(\dot H+H^2)$ and
$R=6(\dot H+2H^2)$. Varying the lapse $N$ of $\dd s^2=-N^2\dd t_{\rm c}^2+a^2\dd\vec x^2$
before setting $N=1$ gives the standard $f(R,G)$ constraint
\cite{DeFelice:2009ak}
\begin{equation}
 E_0\equiv-L+GL_G+6(\dot H+H^2)L_R-6H\dot L_R-24H^3\dot L_G=0 ,
 \label{eq:E0}
\end{equation}
where $L_R=\partial L/\partial R$ and $L_G=\partial L/\partial G$ include the
curvature dependence of the couplings, and dots are derivatives with respect to
the cosmic time $t_{\rm c}$; the unsubscripted $t$ remains the RG time, so that
$\dd/\dd R=(2R)^{-1}\dd/\dd t$ below.
The scale-factor equation follows from Eq.~\eqref{eq:E0} by the diffeomorphism
identity. On dS, $\dot H=0$, $R=12H^2$, $G=R^2/6$ and $\dot L_R=\dot L_G=0$.
With $L=-R^2/\xi(R)+\sigma(R)G$, $\dd/\dd R=(2R)^{-1}\dd/\dd t$, one has
$L_G=\sigma$ and $L_R=-2R/\xi+R\beta_\xi/2\xi^2+G\beta_\sigma/2R$, so that
\begin{equation}
 E_0\big|_{\rm dS}=\frac{R^2}{\xi}-\sigma G+\sigma G
 +\frac{R}{2}\Big[-\frac{2R}{\xi}+\frac{R\beta_\xi}{2\xi^2}+\frac{G\beta_\sigma}{2R}\Big]
 =\frac{R^2}{4}\Big(\frac{\beta_\xi}{\xi^2}+\frac{\beta_\sigma}{6}\Big),
\end{equation}
which is Eq.~\eqref{eq:dScond}. The same combination is the coefficient of $R^2$ in the
beta-function contribution to the trace,
$\Theta_\beta=(\beta_\xi/\xi^2)R^2+(\beta_\lambda/2\lambda^2)C^2+\beta_\sigma E_4+b_{\Box R}\Box R$,
evaluated on dS ($C^2=\Box R=0$, $E_4=R^2/6$); $b_{\Box R}$ is scheme dependent
and does not contribute there \cite{Duff:1993wm,Einhorn:2014gfa}.

\section{Euclidean action, boundary terms and the invariant \texorpdfstring{$C_E$}{C\_E}}
\label{app:euclid}

On the round four-sphere of radius $a$, $R=12/a^2$, $E_4=R^2/6=24/a^4$ and
${\rm Vol}(S^4)=8\pi^2a^4/3$, so $\int_{S^4}\sqrt g\,R^2=384\pi^2$ and
$\int_{S^4}\sqrt g\,E_4=64\pi^2=32\pi^2\chi(S^4)$. For the hemisphere both
integrals are halved, and the Euclidean action $I_E=-iS$ of Eq.~\eqref{eq:action} is
\begin{equation}
 I_{E,1/2}(t)=\frac{192\pi^2}{\xi(t)}-32\pi^2\sigma(t)=-192\pi^2h(t)
 =-\frac{192\pi^2}{\xi'}-32\pi^2C_E,\qquad
 h=\frac{\sigma}{6}-\frac1\xi=\frac{C_E}{6}+\frac{1}{\xi'} .
\end{equation}
For a compact manifold with boundary the Euler integral is completed by the
Chern--Simons-like boundary term $Q_3$,
$\int_M\sqrt g\,E_4+\int_{\partial M}\sqrt q\,Q_3=32\pi^2\chi(M)$ ($q$ the
induced metric); at the
totally geodesic equator both $Q_3$ and the generalized Gibbons--Hawking term
$f_RK$ ($f=-R^2/\xi$, $K$ the trace of the extrinsic curvature) vanish in value, although
they are part of the off-shell variational
problem \cite{Myers:1987yn,Dyer:2008hb}. Varying the radius of the full sphere
avoids changing boundary data and gives the same stationarity condition
$\dd h/\dd t=0$; explicitly, $I_E(S^4)=384\pi^2/\xi-64\pi^2\sigma$ and
\begin{equation}
 \frac{\dd I_E(S^4)}{\dd t}=-384\pi^2\frac{\beta_\xi}{\xi^2}-4\NGB ,
 \label{eq:sphere-derivative}
\end{equation}
where the second term is the integrated Euler ($a$-type) trace anomaly,
$\int_{S^4}\sqrt g\,\langle T^\mu{}_\mu\rangle\supset-4\NGB$ in our
normalization, which follows from the Weyl Ward identity and is therefore
independent of the RG prescription used for $\beta_\xi$ \cite{Shinn:2026pc}.
Setting Eq.~\eqref{eq:sphere-derivative} to zero reproduces Eq.~\eqref{eq:dScond}. With the
momentum-induced $\beta_\xi$ of Table~\ref{tab:beta} instead, the same
condition reads $(1-6\NGB)(\xi/\lambda)^2-36\,\xi/\lambda-2520=0$, whose
discriminant $11376-60480\,\NGB$ is negative for $\NGB>79/420$; since the
graviton contribution alone gives $\NGB\ge196/45$, that flow admits no
stationary round sphere, and in particular the sphere of Ref.~\cite{Liu:2025qg}
at $\beta_\xi=0$ is not stationary \cite{Shinn:2026pc}. In the standard flow
the condition $(6\NGB-5)(\xi/\lambda)^2+180\,\xi/\lambda-360=0$ has one
negative and one positive root for every $\NGB>5/6$; we use the negative one. The radius Hessian of the hemisphere action at the saddle is
$I_{E,1/2,tt}|_d=-192\pi^2h_{tt}|_d<0$ (twice this for the full sphere) since $h$ has a strict minimum there
(Appendix~\ref{sec:flow}); this restricted negative direction does not by itself
fix the fluctuation spectrum or the integration contour of a no-boundary
amplitude \cite{Hartle:1983ai,Hawking:1984ph,Lehners:2023yrj,Jonas:2020dsf,Maldacena:2024uhs}.

\begin{figure}[t]
 \centering
 \includegraphics[width=0.98\linewidth]{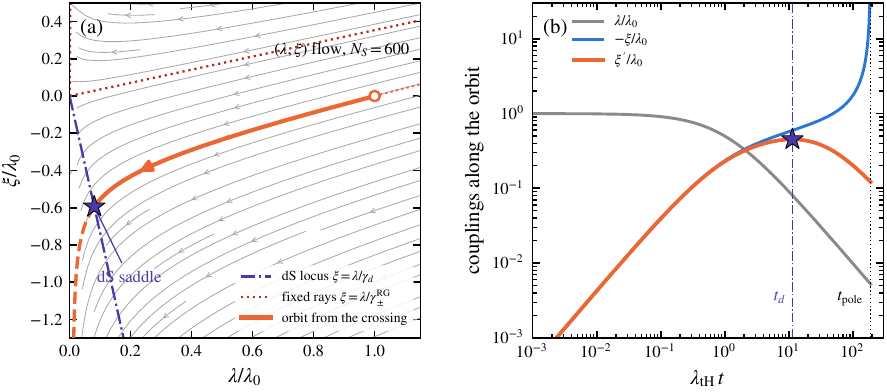}
 \caption{(a) Phase portrait of the closed $(\lambda,\xi)$ one-loop flow for
 $N_S=600$ in units of the crossing coupling $\lambda_0$ (grey streamlines,
 arrows toward increasing $\mu$), with the negative-$\xi$ dS locus
 $\xi=\lambda/\gamma_d$ (dash-dotted), the fixed rays
 $\xi=\lambda/\gamma^{\rm RG}_\pm$ (dotted) and the orbit from the crossing
 (orange; open circle at $t=0$, star at the saddle, dashed beyond it, dotted
 for the formal positive-$\xi$ continuation). (b) $\lambda$, $-\xi$ and the
 Euler-completed $\xi'$ of Eq.~\eqref{eq:hCE} along the same orbit, against
 $\lth t$: $\xi'$ is maximal at $t_d$ (star) although $\xi$ is monotonic;
 $-\xi$ diverges at the Riccati pole.}
 \label{fig:flow2d}
\end{figure}

The combination $C_E=\sigma-\NGB/(\Ncal\lambda)$ is exactly conserved by Eq.~\eqref{eq:betas},
since $\dd[\NGB/(\Ncal\lambda)]/\dd t=-\NGB\beta_\lambda/(\Ncal\lambda^2)=\NGB/16\pi^2=\beta_\sigma$.
It is the additive integration constant of the Euler coupling,
$C_E=\sigma_0-\NGB/(\Ncal\lambda_0)$, and $h=C_E/6+1/\xi'$ with the
$C_E$-independent Euler-completed coupling $1/\xi'=\NGB/(6\Ncal\lambda)-1/\xi$
of Eq.~\eqref{eq:hCE}. A shift $\delta C_E$ at fixed $(\lambda,\xi)$ changes the hemisphere
action by $-32\pi^2\delta C_E$, a common factor for all geometries of fixed
topology that does not move any saddle; relative weights of different
topologies do change. Since $\beta_\lambda$ and $\beta_\xi$ do not depend on
$\sigma$, the full three-dimensional flow is the closed $(\lambda,\xi)$ flow
times a uniform drift $\dd\sigma/\dd t=\NGB/16\pi^2$; the surfaces of constant
$C_E$ foliate it and are translates of one another. In the coordinates
$(\lambda,\xi')$ the flow is the polynomial vector field
\begin{equation}
 16\pi^2\beta_{\xi'}=-\xi'^2\,\frac{D_{\rm dS}(\gamma)}{36},\qquad
 \gamma=\frac{\NGB}{6\Ncal}-\frac{\lambda}{\xi'},
\end{equation}
which is what is drawn in Fig.~\ref{fig:flow}(a). The negative-$\xi$ branch occupies the
wedge $0<\xi'<6\Ncal\lambda/\NGB$, whose upper edge is $\xi\to-\infty$ (the
Riccati pole), not a regular continuation. Figure~\ref{fig:flow2d} shows the
phase portrait in the original couplings, where the dS locus is the ray
$\xi=\lambda/\gamma_d$ crossed once by the orbit, and the couplings along the
orbit: $\xi$ is monotonic and diverges at the pole, while the Euler-completed
$\xi'$ has its maximum at $t_d$.

\section{Exact solution of the flow and location of the saddle}
\label{sec:flow}

With $u\equiv\xi/\lambda$ and $\dd\tau\equiv\lambda\,\dd t/16\pi^2$, Eq.~\eqref{eq:betas}
gives the autonomous Riccati equation
$\dd u/\dd\tau=-10+(\Ncal+5)u-\frac{5}{36}u^2=-\frac{5}{36}(u-u_+)(u-u_-)$,
with $u_\pm=1/\gamma_\mp^{\rm RG}$ and
$\gamma_\pm^{\rm RG}=[\Ncal+5\pm\sqrt{(\Ncal+5)^2-50/9}]/20$. Since
$\lambda=\lambda_0/(1+\lth t)$, $\tau=\Ncal^{-1}\ln(1+\lth t)$, and the solution
with $u(0)=0$ is
\begin{equation}
 u(t)=\frac{w-1}{\gamma_-^{\rm RG}w-\gamma_+^{\rm RG}},\qquad
 w=(1+\lth t)^\nu,\qquad \nu=\frac{\sqrt{(\Ncal+5)^2-50/9}}{\Ncal},
\end{equation}
which is Eq.~\eqref{eq:orbit}. The ratio $\gamma=\lambda/\xi=1/u$ runs monotonically from
$-\infty$ at $t=0^+$ to $0^-$ at the pole $w=\gamma_+^{\rm RG}/\gamma_-^{\rm RG}$,
with $16\pi^2\beta_\gamma=\lambda[10\gamma^2-(\Ncal+5)\gamma+5/36]>0$ for
$\gamma<0$. The dS condition $D_{\rm dS}(\gamma)=0$ is met once, at
$w_d=(\gamma_d-\gamma_+^{\rm RG})/(\gamma_d-\gamma_-^{\rm RG})$, giving Eq.~\eqref{eq:saddle}.
Because $\dd h/\dd t=D_{\rm dS}(\gamma)/(36\cdot16\pi^2)$ and
$D_{\rm dS}'(\gamma_d)=180(1-4\gamma_d)>0$, $h$ has a strict minimum and
$\xi'=(h-C_E/6)^{-1}$ a strict maximum at $t_d$ whenever $\xi'_d$ is finite; explicitly
\begin{equation}
 h_{tt}\big|_d=\frac{5(1-4\gamma_d)}{16\pi^2}\,\beta_{\gamma,d},\qquad
 \beta_{\gamma,d}=\frac{\lambda_d}{16\pi^2}\Big(\frac{\NGB}{6}-\Ncal\gamma_d\Big),\qquad
 \frac{\dd^2\xi'}{\dd t^2}\Big|_d=-\xi'^2_d\,h_{tt}\big|_d<0 .
\end{equation}
For the illustrative $N_S=600$ flow ($\Ncal=23.3$, $\NGB=6.0222$):
$\gamma_d=-0.13598$, $\gamma_\pm^{\rm RG}=(2.8251,\,0.004916)$, $\nu=1.2104$,
$\lambda_d/\lambda_0=0.08078$, $\xi'_d/\lambda_0=0.4512$, $\lth t_d=11.38$ and
$\lth t_{\rm pole}=189.5$. For the CMB-normalized trajectory
($\Ncal=3.5026\times10^5$, $\NGB=5.8379\times10^4$, $\lambda_0=3.6533\times10^{-4}$):
$\gamma_d=-30.943$, $\lambda_d=3.2249\times10^{-7}$, $\xi_d=-1.0422\times10^{-8}$,
$t_d=1396.8$ and $t_{\rm pole}=1.09\times10^{11}$. We verified Eq.~\eqref{eq:orbit} against
direct numerical integration of Eq.~\eqref{eq:betas} to a relative accuracy of $10^{-12}$
(script \texttt{check\_numbers.py}).

\section{Scalaron description}
\label{app:scalaron}
\label{sec:scalaron}

\subsection{Auxiliary fields and the Euler remainder}

In the conformally flat sector, $L(R,G)=f(R)+\sigma(R)G$ with
$f=-R^2/\xi(R)$ has Hessian determinant $-\sigma_R^2$ with respect to $(R,G)$;
where $\sigma_R\neq0$ an equivalent representation uses two auxiliary fields,
$S_{\rm aux}=\int\sqrt{-g}\,[PR+QG-U(P,Q)]$ with $U=\rho(Q)P-f(\rho(Q))$ and
$\rho$ the local inverse of $\sigma$, so that $P=L_R$ and $Q=L_G$ on shell.
For $P>0$ the conformal transformation yields the algebraic potential
$\mpl^4U/4P^2$ together with a field-dependent Euler coupling and derivative
interactions \cite{DeFelice:2009ak}; the potential alone is not a complete
single-field description. Two regimes allow the scalaron potential to be
isolated: near dS (covariant matching, below) and in the small-ratio rolling
regime (Appendix~\ref{sec:roll}).

Subtracting $\sigma_d=\sigma(R_d)$ and defining
$c_{\rm p}=-1/\xi+(\sigma-\sigma_d)/6$,
\begin{equation}
 -\frac{R^2}{\xi}+\sigma E_4=R^2c_{\rm p}+\sigma_dE_4+(\sigma-\sigma_d)\Big(E_4-\frac{R^2}{6}\Big),\qquad
 E_4-\frac{R^2}{6}=C^2-2S_{\mu\nu}S^{\mu\nu},\quad S_{\mu\nu}=R_{\mu\nu}-\tfrac14Rg_{\mu\nu}.
\end{equation}
The constant $\sigma_dE_4$ has no bulk variation after boundary completion.
Both $C_{\mu\nu\rho\sigma}$ and $S_{\mu\nu}$ vanish on the maximally symmetric
background and are first order in the metric perturbation, while
$\sigma-\sigma_d=\NGB(t-t_d)/16\pi^2$ is first order in $\delta R$; the
remainder is therefore cubic and $f_{\rm p}(R)=R^2c_{\rm p}(R)$, with the
unchanged Weyl sector, reproduces the theory through quadratic order about dS.
On flat FLRW, $E_4-R^2/6=-6\dot H^2$, second order in slow roll; the
smallness of the remainder in the action does not by itself bound its
variation when the leading scalar force is small, which is why the rolling
regime is treated separately in Appendix~\ref{sec:roll} and the Euler correction
estimated in Appendix~\ref{sec:corrections}.

\subsection{Hilltop mass}

\begin{figure}[t]
 \centering
 \includegraphics[width=0.5\linewidth]{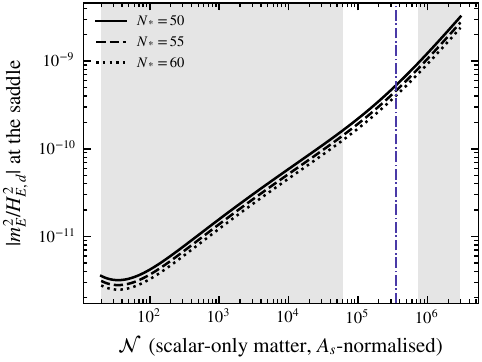}
 \caption{Tachyonic mass of the scalaron at the dS saddle, Eq.~\eqref{eq:hilltop}, along
 trajectories normalized to $A_s=2.1\times10^{-9}$ with scalar-only matter, as
 a function of the total coefficient $\Ncal$ for $N_*=50,55,60$. Grey bands
 are excluded by $r<0.034$ (left) and by strong coupling before the end of
 inflation (right); the dash-dotted line is the representative solution.}
 \label{fig:hilltop}
\end{figure}

Write $c\equiv c_{\rm p}$, $x=\ln(R/R_d)$, primes for $\dd/\dd x$, so that
$c'=\frac12(\beta_\xi/\xi^2+\beta_\sigma/6)$ vanishes at the saddle and
$c''|_d=\frac14\,\dd^2c_{\rm p}/\dd t^2=\frac{5\lambda_d(1-4\gamma_d)(\NGB/6-\Ncal\gamma_d)}{4(16\pi^2)^2}>0$.
From $V=\frac{\mpl^4}{4}(c+c')/(2c+c')^2$ and $\varphi=\sqrt{3/2}\,\mpl\ln[2R(2c+c')/\mpl^2]$,
\begin{equation}
 V'\big|_d=0,\qquad
 V''\big|_d=-\frac{\mpl^4c''}{16c^2}\Big(1+\frac{c''}{2c}\Big),\qquad
 \varphi'\big|_d=\sqrt{\tfrac32}\,\mpl\Big(1+\frac{c''}{2c}\Big),\qquad
 V_d=\frac{\mpl^4}{16c_d},
\end{equation}
so that $m_E^2=V''/\varphi'^2=-\mpl^2c''/[24c^2(1+c''/2c)]$ and, with
$H_{E,d}^2=V_d/3\mpl^2=\mpl^2/48c_d$,
\begin{equation}
 \frac{m_E^2}{H_{E,d}^2}=-\frac{4c''}{2c+c''}=-4\Big[1+\frac{2c_d}{c''_d}\Big]^{-1},\qquad
 c_d=-\frac{\gamma_d}{\lambda_d},
\end{equation}
which is Eq.~\eqref{eq:hilltop}. The Jordan-frame ratio $m_J^2/H_{J,d}^2$ is identical since
both scale with the same conformal factor. In the weak-running limit
$c''\ll c$, $|m_E^2/H_{E,d}^2|\simeq2c''/c\simeq\sqrt{5\NGB/3\Ncal}\,g_d^2/\sqrt\Ncal$
with $g_d=\Ncal\lambda_d/16\pi^2$ and $\NGB/\Ncal$ fixed. Figure~\ref{fig:hilltop}
shows $|m_E^2/H_{E,d}^2|$ along $A_s$-normalized trajectories as a function of
$\Ncal$: it lies between $10^{-10}$ and $10^{-9}$ across the allowed window, so
the saddle is always an extremely flat maximum. Because $V_{\rm p}/V_d-1$ is of
order $10^{-14}$ near the saddle, double-precision finite differences are
unreliable there; we verified Eq.~\eqref{eq:hilltop} in 60-digit arithmetic against the
canonical second derivative $(V_{tt}-V_{,\varphi}\varphi_{tt})/\varphi_t^2$ of
$V_{\rm p}$, with step sizes $10^{-2}$--$10^{-5}$ in $t$ converging to the
analytic value $m_E^2/H_{E,d}^2=-4.5608444447\times10^{-10}$ for the
representative trajectory to better than twelve digits
(\texttt{check\_numbers.py}).

\section{Reduced flow, potential and slow-roll algebra}
\label{app:roll}\label{sec:roll}

Dropping the $\lambda\xi$ and $\xi^2$ terms of $\beta_\xi$, the flow
$16\pi^2\beta_\lambda=-\Ncal\lambda^2$, $16\pi^2\beta_\xi=A\lambda^2$ with
$\xi(0)=0$ integrates to Eq.~\eqref{eq:reduced}. For the standard prescription $A=-10$ and
$\xi<0$, so $f(R)=-R^2/\xi>0$; a comparison with $A>0$ requires the
corresponding sign convention for the action. With $\kappa\equiv16\pi^2/(|A|\lambda_0^2)$,
$R=R_0e^{2t}$ and $D=t(1+\lth t)$, using $\dd/\dd t[(1+\lth t)/t]=-1/t^2$,
\begin{equation}
 f=\kappa R^2\frac{D}{t^2},\qquad
 F\equiv f_R=\frac{1}{2R}\frac{\dd f}{\dd t}=\kappa R\,\frac{4D-1}{2t^2},\qquad
 RF-f=\kappa R^2\,\frac{2D-1}{2t^2},
\end{equation}
so that $V=\mpl^4(RF-f)/4F^2$ and $\varphi=\sqrt{3/2}\,\mpl\ln(2F/\mpl^2)$ give
Eq.~\eqref{eq:Vroll}. The chart requires $F>0$, i.e.\ $D>1/4$, and $V>0$ requires $D>1/2$.
Differentiating along $t$, with $\dd D/\dd t=1+2\lth t$ and $t\,\dd D/\dd t=2D-t$,
\begin{equation}
 \frac{\dd\varphi}{\dd t}=\sqrt{\tfrac32}\,\mpl\,\frac{2(4tD-3t+1)}{t(4D-1)},\qquad
 \frac{\dd\ln V}{\dd t}=\frac{2(4tD-3t+1)}{t(2D-1)(4D-1)},
\end{equation}
whence $\mpl V_{,\varphi}/V=\sqrt{2/3}/(2D-1)$, $\epsilon_V=1/[3(2D-1)^2]$ and
$r=16\epsilon_V$. A second differentiation gives
\begin{equation}
 \eta_V=\mpl^2\frac{V_{,\varphi\varphi}}{V}=\frac{2}{3(2D-1)^2}-\frac{2(2D-t)(4D-1)}{3(2D-1)^2(4tD-3t+1)},
\end{equation}
and $n_s=1-6\epsilon_V+2\eta_V$ reproduces Eq.~\eqref{eq:nsr} after using
$4D^2-3D+1+\lth t=(1+\lth t)(4tD-3t+1)$. The amplitude
$A_s=V/(24\pi^2\mpl^4\epsilon_V)$ is Eq.~\eqref{eq:As}. The $e$-fold measure
$\dd N=[V/(\mpl^2V_{,\varphi})]\,\dd\varphi$ becomes
\begin{equation}
 \frac{\dd N}{\dd t}=\frac{3(2D-1)(4tD-3t+1)}{t(4D-1)}
 =6\lth t^2+6t-6+\frac3t-\frac{3(1+2\lth t)}{4D-1},
\end{equation}
whose primitive is $I(t)$ of Eq.~\eqref{eq:Nefold}. Inflation ends at $\epsilon_V=1$,
$2D_e-1=1/\sqrt3$, i.e.\ $t_e=(1+1/\sqrt3)/[1+\sqrt{1+2\lth(1+1/\sqrt3)}]$;
then $0<t_e<t_*$ and the canonical chart reaches $D=1/4$ before the formal
crossing when followed toward decreasing curvature. All of these expressions
were checked against finite-difference derivatives of $V(\varphi(t))$ and direct
quadrature of $\dd N$ (\texttt{check\_numbers.py}).

For $\lth t\gg1$ and $D\gg1$,
\begin{equation}
 \frac{\varphi-\bar\varphi_0}{\mpl}\simeq\sqrt6\,t,\qquad
 \bar\varphi_0\equiv\sqrt{\tfrac32}\,\mpl\ln\frac{64\pi^2\lth R_0}{|A|\lambda_0^2\mpl^2},\qquad
 V\simeq V_0\Big(1-\frac{1}{\lth t}\Big),
\end{equation}
with $V_0=\mpl^4|A|\lambda_0/16\Ncal$, which is Eq.~\eqref{eq:plateau}. Then $\epsilon_V\simeq1/(12\lth^2t^4)$, $N\simeq2\lth t^3$, and Eq.~\eqref{eq:limits} follows;
the two forms of $r$ in Eq.~\eqref{eq:limits} correspond to holding $\lth$ or $\Ncal$ fixed.

\section{Comparison with data: details}

\begin{table}[t]
 \centering
 \caption{Landmarks of the prediction curves of Fig.~\ref{fig:nsr} for
 $A_s=2.1\times10^{-9}$. ``Pure gravity'' is $\Ncal=133/10$; the tensor bound
 is $r<0.034$ (95\%, Ref.~\cite{Balkenhol:2025inf}, with or without DESI); the
 strong-coupling limit is $g(t_e)=\Ncal\lambda(t_e)/16\pi^2=1$, which occurs at
 $\lth=1.789$ independently of $N_*$. The last three columns give the
 approximate $\chi^2$ of the strong-coupling-limit point for the three tilt
 measurements (``+DESI'' denotes Planck--SPT--ACT--BK18 with DESI), with the half-Gaussian tensor term.}
 \label{tab:landmarks}
 \setlength{\tabcolsep}{3pt}
 \footnotesize
 \begin{tabular}{c ccc ccc ccc ccc}
 \toprule
 & \multicolumn{3}{c}{pure gravity} & \multicolumn{3}{c}{$r=0.034$} & \multicolumn{3}{c}{$g(t_e)=1$} & \multicolumn{3}{c}{$\chi^2$ at $g(t_e)=1$}\\
 $N_*$ & $10^5\lth$ & $n_s$ & $r$ & $\lth$ & $10^{-4}\Ncal$ & $n_s$ & $r_{\min}$ & $n_s$ & $10^{-5}\Ncal$ & P-ACT-LB & +DESI & SPA+BK\\
 \midrule
 50 & 2.78 & 0.9727 & 0.0680 & 0.193 & 8.8 & 0.9730 & 0.0105 & 0.9736 & 6.6 & 0.41 & 0.44 & 3.2\\
 55 & 2.60 & 0.9751 & 0.0622 & 0.148 & 7.3 & 0.9754 & 0.0093 & 0.9760 & 7.0 & 0.53 & 1.47 & 6.2\\
 60 & 2.45 & 0.9771 & 0.0573 & 0.114 & 6.0 & 0.9773 & 0.0084 & 0.9779 & 7.4 & 1.38 & 3.38 & 9.5\\
 \bottomrule
 \end{tabular}
\end{table}

The measurements used are the scalar tilt of the ACT DR6 combination with
Planck, CMB lensing and DESI BAO (P-ACT-LB), $n_s=0.9743\pm0.0034$
\cite{ACT:2025lcdm,ACT:2025extended}, and the two combinations of
Ref.~\cite{Balkenhol:2025inf}: Planck--SPT--ACT--BK18 with DESI,
$n_s=0.9728\pm0.0029$, and without DESI, $n_s=0.9682\pm0.0032$; both give the
95\% limit $r<0.034$, which DESI leaves unchanged. We approximate the
marginalized posteriors by a Gaussian in $n_s$ and a half-Gaussian in $r\ge0$
whose 95\% quantile is that limit, $P(r<0.034)={\rm erf}[0.034/(\sqrt2\sigma_r)]=0.95$,
i.e.\ $\sigma_r=0.034/1.960=0.0173$; $\chi^2=[(n_s-\bar n_s)/\sigma_n]^2+(r/\sigma_r)^2$
and the 68\% and 95\% regions of Fig.~\ref{fig:nsr}(a) are $\Delta\chi^2=2.28$ and $5.99$,
the two-parameter levels (the truncation at $r=0$ removes the same fraction of
every $\Delta\chi^2$ shell, so these levels are unchanged). This ignores the correlation between
$n_s$ and $r$ and the non-Gaussian shape of the tensor posterior; the
published marginalized posterior of BK18 with Planck and BAO peaks near
$r\simeq0.014$ \cite{BICEP:2021xfz,Tristram:2021tvh}, which would slightly
favour the upper part of the allowed window. The model curves are computed
from Eqs.~\eqref{eq:nsr}--\eqref{eq:Nefold}: for each $N_*$ and $\lth$ we solve $N_*=I(t_*)-I(t_e)$
for $t_*$ by bracketing, evaluate $(n_s,r)$, and obtain $\Ncal$ from Eq.~\eqref{eq:As}.
Table~\ref{tab:landmarks} lists the landmarks quoted in the main text. The
loop-expansion parameter $g(t)=\Ncal\lambda(t)/16\pi^2=\lth/(1+\lth t)$ decreases
with $t$, so along the inflationary trajectory it is largest at the end of
inflation; we therefore impose one-loop control there, $g(t_e)<1$, which gives
$\lth<1.789$. Imposing the same bound on $\lth=g(0)$ instead would refer to the
formal crossing, which no cosmological solution reaches, and is stronger by the
same factor: it would raise $r_{\min}$ to $0.0147$, $0.0131$ and $0.0118$ for
$N_*=50,55,60$ (at $\Ncal=(4.0$--$4.5)\times10^5$), leaving the qualitative
picture unchanged. Figure~3(b)
uses the complementary parametrization of the earlier analysis
\cite{Liu:2025qg}: for each point of the $(\lth,\Ncal)$ plane, $t_*$ is fixed
by Eq.~\eqref{eq:As} and $N_*$ is derived, so the contours are regions rather than
curves and no restriction on $N_*$ is imposed; the lines $N_*=50,55,60$ are
overlaid to locate the curves of Fig.~\ref{fig:nsr}(a). Along a line of fixed $N_*$ the
two parametrizations coincide. The
representative solution has $N_*=55$, $\lth=0.810327$, $t_*=2.949676$,
$t_e=0.546585$, $D_*=10.000$, $\Ncal=3.5026\times10^5$,
$\lambda_0=3.65333\times10^{-4}$, $n_s=0.975854$, $r=0.0147738$,
$g(t_*)=0.239$, $g(t_e)=0.562$, and $\chi^2=0.93$, $1.83$, $6.45$ for the three
tilt measurements. Its exact full-flow crossing coupling (the $\lambda_0$
that reproduces $A_s$ with Eq.~\eqref{eq:orbit} instead of Eq.~\eqref{eq:reduced}) differs from the
reduced value by $7\times10^{-6}$ relatively.

\section{Euler and Weyl corrections in the observable interval}
\label{app:corrections}\label{sec:corrections}

\paragraph*{Euler term.} Project the running Euler coupling onto an
Einstein-scalar--Gauss--Bonnet form,
$\mathcal L\supset-\frac12\Xi(\varphi)E_4$ with $\Xi=-2\sigma$. In reduced Planck
units the slow-roll scalar equation is $3H\dot\varphi\simeq-V_{,\varphi}-12\Xi_{,\varphi}H^4$
\cite{Guo:2010jr}; with $H^2\simeq V/3$ this is
$\dd\varphi/\dd\ln a\simeq-L_\varphi(1+d)$, $L_\varphi=V_{,\varphi}/V$, $d=4\Xi_{,\varphi}V/3L_\varphi$.
On the reduced solution $\Xi_{,\varphi}=-2(\NGB/16\pi^2)\,\dd t/\dd\varphi$, and using
$\dd\varphi/\dd t$, $V$ and $L_\varphi$ from Appendix~\ref{sec:roll} together with
Eq.~\eqref{eq:As},
\begin{equation}
 d=-\frac{2\NGB A_s\,t_*(4D_*-1)}{3(2D_*-1)(4t_*D_*-3t_*+1)}
  =-\frac{2\NGB A_sD_*(4D_*-1)}{3(2D_*-1)P_*},\qquad
 \delta_1\equiv4H\dot\Xi\simeq-\frac{r}{8}(1+d)d .
\end{equation}
Here $\delta_1$ is the Gauss--Bonnet slow-roll parameter of Ref.~\cite{Guo:2010jr},
which controls the correction to the tensor spectrum. For the representative solution, the scalar-only and vector-only edges of
$\NGB$ at $\Ncal=3.5\times10^5$ ($5.84\times10^4$ and $3.02\times10^5$) give
$|d|=4.49\times10^{-6}$ and $2.32\times10^{-5}$, $|\delta_1|=8.3\times10^{-9}$
and $4.3\times10^{-8}$. These are diagnostics of the stated projection, not
the full perturbation theory of the running-Euler action. Their smallness is
compatible with the Euler term balancing the curvature response at the saddle:
the balance involves $\beta_\sigma$ against $\beta_\xi/\xi^2\propto\lambda^2/\xi^2$,
which is enormous in the rolling regime and falls to $\NGB$ only at
$t_d\simeq1.4\times10^3$ [Fig.~\ref{fig:flow}(b)].

\paragraph*{Weyl term.} With the Einstein-frame gravitational term
normalized to $\mpl$, the spin-2 mode of $-C^2/2\lambda$ has
$M_2^2=\lambda\mpl^2/2$ in that frame \cite{Salvio:2018crh}, so
\begin{equation}
 \frac{M_2^2}{H_E^2}=\frac{3\lambda\mpl^4}{2V}
 =\frac{3\Ncal}{|A|\lth t}\,\frac{(4D-1)^2}{D(2D-1)},
\end{equation}
which is $3.52\times10^5$ at the representative pivot. In the heavy-mode
treatment the tensor power is rescaled by $(1+2H_E^2/M_2^2)^{-1}$
\cite{Clunan:2009er,Deruelle:2010kf,Salvio:2017xul}, a $5.7\times10^{-6}$
effect. A positive heavy mass does not remove the negative residue of the
extra pole; its interpretation depends on the quantization
\cite{Donoghue:2018izj,Anselmi:2020lpp}.

\section{Comparison with momentum-induced beta functions}
\label{app:compare}

\begin{table}[t]
 \centering
 \caption{The two one-loop prescriptions in the original couplings. In each
 column $\Ncal$ denotes that prescription's total coefficient multiplying
 $-\lambda^2/16\pi^2$; at fixed species counts the momentum-induced value
 exceeds the standard one by $14/3$. Dropping the terms containing $\xi/\lambda$
 gives the common reduced form of Eq.~\eqref{eq:reduced} with different signed $A$.}
 \label{tab:beta}
 \setlength{\tabcolsep}{10pt}
 \begin{tabular}{lll}
 \toprule
 & Standard local \cite{Fradkin:1981iu,Avramidi:1985ki,Salvio:2014soa} & Momentum-induced \cite{Buccio:2024hys,Liu:2025qg} \\
 \midrule
 $16\pi^2\beta_\lambda/\lambda^2$ & $-\Ncal$ & $-\Ncal+2\xi/9\lambda$ \\
 $16\pi^2\beta_\xi/\lambda^2$ & $-10+5\xi/\lambda-5(\xi/\lambda)^2/36$ & $70+\xi/\lambda-(\xi/\lambda)^2/36$ \\
 $A$ & $-10$ & $70$ \\
 gravitational part of $\Ncal$ & $133/10$ & $1617/90$ \\
 Euler beta function & $\NGB/16\pi^2$ & not supplied \\
 \bottomrule
 \end{tabular}
\end{table}

Table~\ref{tab:beta} compares the standard local one-loop flow used here with
the momentum-induced flow of Ref.~\cite{Buccio:2024hys} used in
Ref.~\cite{Liu:2025qg}; the scheme and gauge dependence of the two
prescriptions is discussed in Refs.~\cite{Kawai:2024aim,Buccio:2025gauge}.
Both reduce to Eq.~\eqref{eq:reduced} in the small-ratio regime, with $A=-10$ and $A=70$
respectively, and the leading observables then depend on the prescription
only through the combination $A_s\Ncal^2/|A|$ in Eq.~\eqref{eq:As}: at fixed
$(N_*,\lth)$ the predicted $(n_s,r)$ are identical and the required
coefficient scales as $\Ncal\propto\sqrt{|A|}$, i.e.\ the momentum-induced
prescription needs $\sqrt7\simeq2.6$ times more matter fields for the same
tensor amplitude [Fig.~\ref{fig:nsr}(b)]. This is a local statement
about the rolling regime. The global theories differ: for the standard action
$A=-10$ and $\xi<0$, whereas the positive-$\xi$ construction of
Ref.~\cite{Liu:2025qg} requires its own scalaron sign convention; matching
complete spectra also requires matching the spin-2 prescription and the
omitted interactions; and no Euler beta function accompanies the quoted
momentum-induced running. On the round sphere this last point does not matter:
the Euler contribution to the scale derivative is the scheme-independent
$a$-anomaly, so Eq.~\eqref{eq:sphere-derivative} applies to both flows, and
the momentum-induced one has no stationary round sphere at all
\cite{Shinn:2026pc}. Within the standard prescription, the maximum of the
scalar-curvature coupling that was taken to select the four-sphere in the
Euler-truncated flow of Ref.~\cite{Liu:2025qg} is replaced by the maximum of
the Euler-completed coupling $\xi'$.

Table~\ref{tab:compare} summarizes what changes between the two papers and
what does not.

\begin{table}[t]
 \centering
 \caption{The scenario of Ref.~\cite{Liu:2025qg} and the present one.
 Quantities quoted for Ref.~\cite{Liu:2025qg} are taken from that paper; the
 last three rows use the same $A_s$ normalization and the strong-coupling
 criterion $g(t_e)<1$ for both.}
 \label{tab:compare}
 \setlength{\tabcolsep}{5pt}
 \footnotesize
 {\raggedcells
 \begin{tabular}{l p{5.2cm} p{5.8cm}}
 \toprule
 & Ref.~\cite{Liu:2025qg} & this work \\
 \midrule
 action, RG scale & Eq.~\eqref{eq:action} without $\sigma E_4$; $\mu^2\propto R$ & Eq.~\eqref{eq:action}; $\mu^2\propto R$ \\
 beta functions & momentum-induced \cite{Buccio:2024hys}, $A=70$ & standard local one-loop, $A=-10$ \\
 Euler (Gauss--Bonnet) running & neglected & retained; $\beta_\sigma=\NGB/16\pi^2$ \\
 UV behaviour on the branch & $\lambda,\xi$ both asymptotically free & $\lambda$ free; $\xi$ has a pole at $t_{\rm pole}\gg t_d$ \\
 early dS state & zero of $\beta_\xi$ (maximum of $\xi$); not stationary once the Euler anomaly is kept \cite{Shinn:2026pc} & Euler balance $6\beta_\xi/\xi^2+\beta_\sigma=0$ (maximum of $\xi'$) \\
 stability of the dS state & unstable through running (qualitative) & hilltop, $m_E^2/H_E^2\simeq-5\times10^{-10}$ (Eq.~\eqref{eq:hilltop}) \\
 rolling potential & inverse-linear plateau & same, Eq.~\eqref{eq:plateau}, with $V_0=\mpl^4|A|\lambda_0/16\Ncal$ \\
 $n_s$, $r$ at fixed $(\lth,N_*)$ & Eqs.~\eqref{eq:nsr}--\eqref{eq:limits} with $|A|=70$ & identical \\
 $r_{\min}$ (strong coupling) & $\simeq0.01$ & $0.0105$, $0.0093$, $0.0084$ ($N_*=50,55,60$) \\
 matter content for given $r$ & $\Ncal\propto\sqrt{70}$ & $\Ncal\propto\sqrt{10}$: $2.6$ times fewer fields \\
 allowed window & $0.1\lesssim\lth\lesssim1$, $10^5\lesssim\Ncal\lesssim10^6$ & $0.11\lesssim\lth\lesssim1.8$, $6\times10^4\lesssim\Ncal\lesssim7\times10^5$ \\
 data & Planck, ACT, SPT, BK, DESI (2024) & P-ACT-LB, Planck--SPT--ACT--BK18(+DESI) (2025) \\
 \bottomrule
 \end{tabular}}
\end{table}

\begin{figure}[t]
 \centering
 \includegraphics[width=0.5\linewidth]{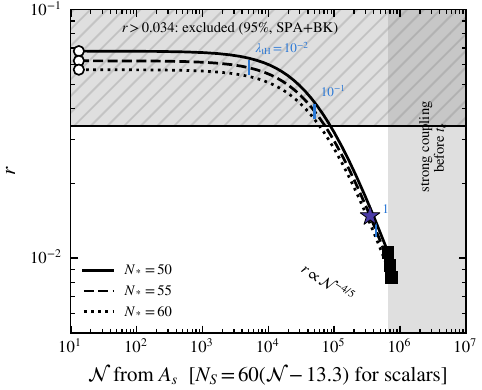}
 \caption{Tensor amplitude against the total coefficient $\Ncal$ fixed by
 $A_s=2.1\times10^{-9}$, for $N_*=50,55,60$ (the curves of Fig.~\ref{fig:nsr}). Open
 circles: pure gravity; squares: the strong-coupling limit $g(t_e)=1$; ticks:
 $\lth$ along the $N_*=55$ curve. Hatched regions are excluded by $r<0.034$ and
 by strong coupling before the end of inflation; in the matter-enhanced regime
 $r\propto\Ncal^{-4/5}$, Eq.~\eqref{eq:limits}.}
 \label{fig:rN}
\end{figure}

\section{Code}

All figures and every number quoted in the main text are produced by four Python
scripts (NumPy, SciPy, Matplotlib), provided as ancillary files with the arXiv submission:
\begin{description}
 \item[\texttt{qgrg.py}] the beta functions, species coefficients, exact orbit
 \eqref{eq:orbit}, saddle \eqref{eq:gammad}--\eqref{eq:saddle}, hilltop \eqref{eq:hilltop}, dS-matched potential \eqref{eq:Vp}, reduced potential
 \eqref{eq:Vroll} and observables \eqref{eq:nsr}--\eqref{eq:Nefold}, and the correction formulae of
 Appendix~\ref{sec:corrections};
 \item[\texttt{check\_numbers.py}] independent numerical checks of every
 analytic result (direct integration of the flow, finite-difference slow-roll
 parameters, quadrature of the $e$-fold integral, high-precision hilltop mass);
 \item[\texttt{fit\_scan.py}] the data model, the scan over $(\lth,N_*)$, and
 the landmarks of Table~\ref{tab:landmarks};
 \item[\texttt{make\_figures.py}] Figs.~\ref{fig:flow}--\ref{fig:nsr} and
 Figs.~\ref{fig:flow2d}, \ref{fig:hilltop} and \ref{fig:rN}.
\end{description}

% ---- bibliography generated by BibTeX (apsrev4-2) from letter.bib; embedded so no BibTeX run is needed
%

\end{document}